# Securing of Unmanned Aerial Systems (UAS) Against Security Threats Using Human Immune System

Reza Fotohi
Faculty of Computer Science and Engineering, Shahid Beheshti University, Tehran, Iran

Corresponding author: Reza Fotohi, fotohi.reza@gmail.com, r_fotohi@sbu.ac.ir

**Abstract**
UASs form a large part of the fighting ability of the advanced military forces. In particular, these systems that carry confidential information are subject to security attacks. Accordingly, an Intrusion Detection System (IDS) has been proposed in the proposed design to protect against the security problems using the human immune system (HIS). The IDSs are used to detect and respond to attempts to compromise the target system. Since the UASs operate in the real world, the testing and validation of these systems with a variety of sensors is confronted with problems. This design is inspired by HIS. In the mapping, insecure signals are equivalent to an antigen that are detected by antibody- based training patterns and removed from the operation cycle. Among the main uses of the proposed design are the quick detection of intrusive signals and quarantining their activity. Moreover, SUAS-HIS method is evaluated here via extensive simulations carried out in NS-3 environment. The simulation results indicate that the UAS network performance metrics are improved in terms of false positive rate, false negative rate, detection rate, and packet delivery rate.

## 1  Introduction

UASs or Aerial vehicles networks (AVNs) are defined as any individual aerial vehicle either in communication with other vehicles, such as a Unmanned Aerial Vehicles (UAVs) with another UAV (U2U), and a UAV with an Satellite (U2S), or are in communication with stationary infrastructures such as an UAV with Traffic control tower (U2T), and a UAV with Ground station (U2G) [1]. A typical scenario for communication among the UAVs is demonstrated in Figure 1. This scenario includes multiple components and utilizes various links for communication. Each link is responsible for transmitting certain types of data and information. In general, according to type of information being transmitted, there should exist three different types of links in these networks, namely radio communication, U2U and Satellite link. In the radio communication links, telemetry data, video, audio, and control information are carried. Moreover, satellite links are responsible for carrying GPS, weather, and meteorological information, in addition to the data carried by the radio communication links. Employing UASs in the networks for ballistic missile defense is one of their applications with the highest level of vitality. In such applications, the tasks for the UAVs are usually patrolling an intermediary land stretched between the site where the ballistic missile is launched, and its intended target. Since ballistic missiles can cruise at an extremely high speeds, they mandate the use of quick detection methods to track and to eliminate. In specific, to increase the chance of successfully intercepting a ballistic missile, it is essential to have swift detection and tracking system, capable of detecting and tracking the missile right after it is launched. The designers working on ballistic missile defense networks are aiming the system to be capable of intercepting the missiles during their initial 2 to 5 minutes of flight, which is called the boost phase. During the boost phase, if the trajectory of the missile is directly away from the trajectory of the UAV, it will easily recede from the range of the sensors on the UAV. Hence, the routing of the information used in the network of the sensors of the ballistic missile should be addressed using networks of hybrid wireless sensors capable of complying with high availability and requirements mandated due to security. To demonstrate the UAVs' applications that have an innate time sensitivity, and to signify the urgency of providing security in communication channels, this case is discussed in the following paper [2, 3]. However, despite the advantage's UAVs provide through various applications, since there are situations where no pilot monitors the activities, they are potentially vulnerable against cyber threats. This intensifies the emergence to design secure and reliable UASs and overcome the challenges to avoid damage and destruction to other systems as well as human lives [2, 3].

A number of attacks, such as Wormhole (WH), Black hole (BH), Gray hole (GH), and Fake Information Dissemination (FID), illegally penetrate the system. Once an unmanned system is affected by an attack, removing the threat and bringing the system back online is a laborious task. It should be mentioned that the common methods to secure information, such as encryption or intrusion detection [4], are inadequate when dealing with such risks. To elaborate, the mentioned schemes do not consider the sensor and actuator measurements compatibility factor with the physical process and control mechanism of the UAV, which are substantial to the protection scheme. In the SUAS-HIS proposed design, the malicious UAV is robust against four lethal attacks (WH, BH, GH, and FID) so that intrusive operations are quickly identified and removed from the spying missions or top-secret information surveillance. The proposed schema also improves critical standards of quality of service including detection rates, false positive rates and false negative rates.

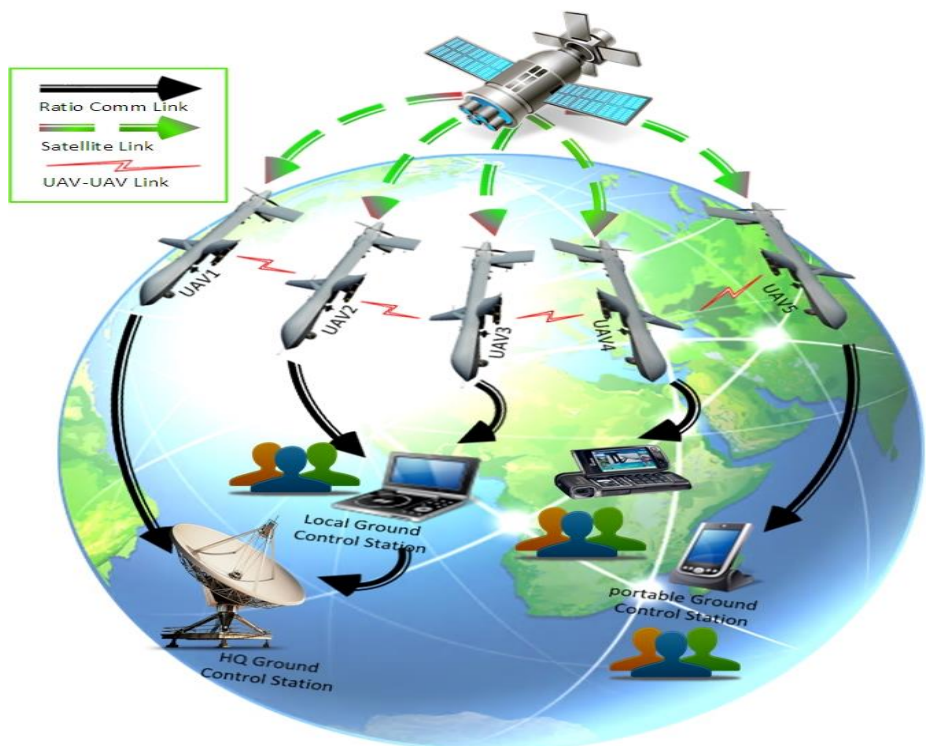

Figure 1: Typical UAS communication scenario.

The paper presented here is organized as the following. Section 2 presents the cyber security threats and detection schemes. Section 3 presents the HIS. In Sect. 4 brings the proposed SUAS-HIS framework. Moreover, parameters utilised for performance evaluation are investigated and simulation results are discussed in Section 5. Finally, in Section 6, the paper is concluded.

## 2 Cyber security threats and detection schemes

In the following section, we are going to discuss the following issues: cyber security threats targeting UAVs, detection schemes to provide protection for the UAS.

### 2.1 Cyber Security Threats

UASs are prone to cyber-security threats and function degradation, be them active threats or passive, due to the fact that they rely on wireless channels for communication. A list of major cyber security threats targeting UAVs is provided in Figure 2. In this paper, the following vulnerabilities are of interest:

- *Wormhole Attack:* WH attacks are a major threatening attack on UAVs. In WH attacks, a hostile node receives data packets at a certain position in UAV, and tunnels the packets to another hostile node at a distant point, which regulates the packets to its neighbouring nodes. Establishing this tunnel is available through multiple methods, such as a channel established out of band, an encapsulated packet, or even a high-powered transmission. In these methods, the packet transmitted through tunnels are received rather promptly or with less hop counts compared to ordinary packets, which are transmitted through a multi-hop route. Through this technique, an illusion is established that the two tunnel end points are close [5]. Therefore, the hostile nodes are established as decoys between the source and destination nodes, and are able to perform subversions such as packet droppings and manipulation.
- *Black Hole Attack:* In this attack, a BH node transmits a forged RREP upon receiving an RREQ packet, claiming shorter and unexpired route, even if the destination entry is missing from the routing table. When the fabricated RREP packet reaches the source node, a route through this malicious intermediate node is established, to discard all legitimate RREP messages transmitted from other intermediate and destination nodes. Therefore, through misleading the source node, BH node successfully attracts the data traffic to that destination. Then, instead of forwarding the incoming messages, the BH node drops all the data packets. When a transmission route is forged, the BH node resets the hop count to a very low value, and the number of destination sequence to a very high value to increase the acceptance chance at the source node. the BH attack can also launch from the source node via fabricating fields-source sequence numbers in RREQ packets and hop counts, resulting in poisoning the routing tables in intermediate nodes as well as the destination nodes [6].
- *Grey Hole Attack:* In grey hole attacks, data transmission in the network is interrupted by malicious nodes via transmitting false routing information. Since the origin of the malicious nodes are unpredictable, grey hole attack is regarded as a BH attack extension. A node may function as both malicious and normal. This attack interrupts route discovery process and causes diminution in throughput and packet delivery ratio [7].
- *FID attack:* An instance of an FID attack is when a fake GPS signal is transmitted by the attacker in order to variate the route of the drone. To do this, the attacker should broadcast interferences between the drone and the pilot, to deteriorate the location estimated for the drone. In studies related to wireless communications [8], there has been various security strategies proposed to overcome an FID attack.

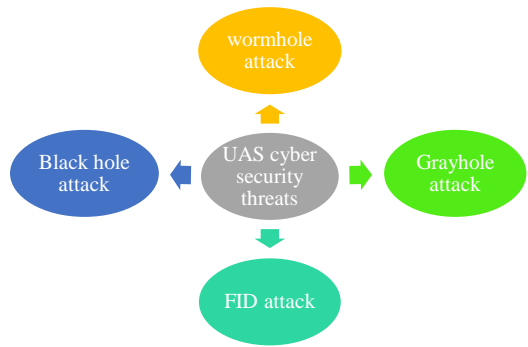

Figure 2: UAS cyber security threats

### 2.2 Detection schemes

The challenges in securing AVNs or UASs using methods to provide cyber detection, and the existing security schemes in literature are discussed in.

Security threats targeting UAV systems are analysed in [9] and a cyber-security threat model is proposed that illustrates plausible attack directions. The model proposed here assists designers and UAV systems operators in understanding the system threat profiles so as to provide addressing various vulnerabilities in the system, identifying high profile threats, and selecting mitigation techniques.

In [10], the authors proposed a security framework to provide protection against malicious behaviour targeting SFA[2] communication systems in aircrafts. As the numerical results presented in this paper indicate, employing the proposed security framework results in detection and prediction rates with high accuracy compared to the intrusion detection methods proposed earlier.

In [11], an adaptive IDS based on device specifications is proposed to detect suspicious UAVs in cooperative operations, where operation continuity is substantial. The proposed IDS system audits the UAVs in a distributed system to determine whether they are functioning normally or if they are under malicious attacks. What they are investigating in this paper is the effectiveness of the proposed rule-based UAV IDS (BRUIDS) behaviour on reckless, random, and opportunistic intrusive behaviours (common cyber-attack behavioural methods). The proposed method bases the audition on behavioural rules to swiftly assess the survivability of a UAV under malicious attacks.

A cooperative IDS based on machine learning is proposed in [12], which is capable of exploiting the data provided from the history efficiently to provide the capability of making decisions proactively. To elaborate, the basis of the model proposed here is employing a de-noising auto encoder (DA) as a building block in order to build a deep neural network. The reason for employing DA is its capability of learning reconstruction of feedbacks of IDS from partial feedbacks. This capability enables us to, in situations where there are no complete feedbacks available from any of the IDSs, make proactive decisions regarding occurrence of any suspicious intrusion.

Rani et al. [13] reviewed the UAV vulnerabilities against attacks. To do so, they provided a general overview over the common hacking methods and the existing defensive, preventive, and trust strategies to avoid damages caused by attacks. To demonstrate the severe consequences of hackings and highlight the significance of developing novel defensive methods, they implemented a hacking procedure on a commercially available UAV named the Parrot AR. Drone. What is demonstrated is that the hacker can cause damages beyond repair and obtain full control over the UAV through compromising the operator/UAV communication link. Moreover, he can regenerate the flight trajectory using Robot Operating System (ROS)–based tools. Studies like this aid in identifying the threats and highlighting the significance of securing these systems against threats.

---

[2] Security framework aircraft

Sedjelmaci et al. [14] proposed a system to protect the UAVs against vital threats; threats targeting data integrity and network availability. The proposed system employs a cyber-detection mechanism to detect deteriorating attacks promptly, upon unfolding. A major issue this paper studied was minimizing the false flags; since classifying a legitimate node may compromise the security system efficiency. Hence, to address the issue, this paper proposed a model to estimate threats based on Belief approach.

In [15], Brust et al. proposed a defensive system for UAVs to intercept and escort a malicious aircraft off the flight zone. The proposed defensive system is a UAV swarm, capable of self-organizing its formation in the event of detecting an intruder, and chasing the malicious UAV as a networked defensive swarm. In this paper, to provide a fully localized approach, the authors utilised the principles of modular design. An innovative auto balancing clustering process is developed to realise the formation based on interception and capture. The simulation results revealed that the resulting networked defensive UAV swarm is flexible against communication losses.

Yoon et al. [16], proposed a solution to network channel or physical hardware hijacking of commercial UAVs by anonymous attackers. The solution proposed here was to exploit an additional encrypted communication channel, an authentication algorithm, and perform DoS attack through Raspberry Pi to maintain UAV control in hijacking situations. The resulting system shows high applicability to commercial UAVs.

Sedjelmaci et al. [17] proposed an IDRF to protect the UAV network against threats targeting integrity of the data and availability of network. To our knowledge, this is the only instance of developing a hybrid-detection technique in UAV networks (i.e., a combination of detection methods based on rules and anomaly detection techniques), while considering the energy constraints of UAV nodes. Simulation results demonstrated in [14] reveals a high attack detection ratio and low false-positive alarms. Plus, the proposed framework requires low communication overhead for a quick response to detected attacks.

In [18], Gao et al. presented a novel distributed algorithm for a team of UAVs to provide an online solution for a self-organization problem in a mission based on search and attack in a hostile environment. This problem requires solving a global optimisation problem, where the proposed distributed SAMSOA separates into a number of local optimisation problems. To do so, each UAV is considered as an individual subsystem and is assigned a dedicated processor, to solve its local optimisation problem. On the other hand, exchanging information in the team enables each subsystem to optimise its decision according to multiple UAVs system. The proposed algorithm considers two separate flight modes, namely normal flight mode and threat avoidance flight mode. In normal mode, the modelled search–attack mission maximises the surveillance coverage ratio, while minimising the existence time for each target. Then, an improved distributed ACO algorithm is designed to generate proper path points. Finally, the path points are connected smoothly by a Dubins curve.

In [19], based on edge matching, a novel approach to detect targets is proposed for UAV formation. A potential function with windowed edges determines the field of attraction for the edges similar to the windowed edges. Then, the problem of detecting the targets based on the edges is formulated as an optimisation problem. Afterwards, to specify location, angle of rotation, and an arbitrary template scale over a specific image, this paper proposed an improved bird swarm algorithm named competitive bird swarm algorithm. To aid the local optimum obtained via the original Bird Swarm Algorithm, converge to global optimal point faster and with more stability, a strategy called "disturbing the local optimum" is proposed. In Table 1, a taxonomy of the common cyber-security threats detection methods developed to protect the UAS against malicious threats is highlighted.

Table 1: Cyber security threats detection schemes for UAS.

| Security schemes | The proposed design performance | Advantages | Disadvantages / constraints |
|---|---|---|---|
| [10] | A thorough investigation on various schemes of cyber detection in AVN is carried out here. Plus, a number of cyber-attacks that are susceptible of occurring is highlighted in this network. Moreover, in order to detect and impede occurring of threats that aim communication systems in aircrafts, a security framework named SFA is proposed in this study. | • high accuracy detection<br>• high prediction rates | • Need for additional hardware<br>• low robustness<br>• high complexity |
| [13] | The design uses the neural network techniques to defend against Wireshark attacks, password theft, Trojan virus and so on to have a safe UAV system | • Full coverage of attacks: password theft, Wireshark, MITM attack, Trojan virus and DDoS attack<br>• Flexibility of the design High intelligence in using neural networks | • Information about all UAVs in the IDS agent<br>• Increased overhead |
| [9] | In this design, various security threats for UAV systems have been analyzed cyber-security threat model have been proposed to detect possible paths affected by the attack. This model uses three important communication components in aircraft carriers. The model also helps designers and users of UAV systems to detect various system threats and apply various techniques to eliminate these types of threats to reduce the effects of these types of attacks. | • Simple architecture<br>• Covering more attacks | • Need for additional hardware<br>• Low accuracy rate |
| [11] | In this design an adaptive specification-based system is used for detecting UAV in which the IDS[3] agent is used in a distributed system to detect whether the UAVs are normal or malicious | • Using Ant Colony Multiple Clustering Model<br>• High accuracy in detecting attacks<br>• Reduced false positive rate<br>• Reduced false negative rate<br>• Continuity of operation | • High overhead in detecting attacks<br>• Not covering more lethal<br>• Need for additional hardware |
| [14] | This paper proposes a cyber-security system to protect UAVs from cyber-attacks that disturb the data integrity and access the network. To face with this lethal attack, an attack estimation model has been proposed based on the belief method | • Has high speed and high accuracy in detecting cyber-attacks, Has the lowest false positive rate, Has the lowest false negative rate | • Not covering other lethal attacks<br>• High energy consumption |
| [15] | The design offers an air defense system to defend malicious UAVs outside the flying area. The proposed UAV defense system includes an air defense capability that can organize itself in the event of intrusion detection and track the malicious UAV as a network shooter | • The use of the modular concept for a completely localized approach<br>• The clustering process with automatic equilibrium to realize the formation of the UAVs and capture their formation<br>• Resistance to intruding UAVs | • Excessive blocking of maneuvers<br>• Random noise generation in three directions that injects the overhead into the network<br>• Not covering most lethal attacks |
| [16] | In this study, a method was developed to prevent the abduction of network channels or physical hardware by unidentified attackers on commercial airless systems. This paper proposes an additional encrypted communication channel, authentication algorithm, and denial of service attack through the likelihood of pi and its high applicability in commercial UAV systems | • Accurate identification of the attackers<br>• Sending data securely for the use of two channels with high reliability | • High overhead operation due to the use of encryption channels<br>• Need for additional hardware<br>• High communication delay |
| [17] | In this design, a set of detection and response techniques is proposed for monitoring UAV behaviors and categorized them into a suitable list (normal, abnormal, suspicious and destructive UAV) according to cyber threats. | • Smart activation of the intrusion monitoring process<br>• High detection rate<br>• Low communication overhead<br>• Rapid intruder detection | • Need for additional hardware |

The previous works to design IDS for the UAS have been listed in Table 2 ("-" indicates the indefinite characteristics).

Table 2: Summary of the approaches for UAS literature.

| References | Placement schema | Detection schema | Attack type | Validation schema |
|---|---|---|---|---|
| [12] | Centralized | Hybrid | Cooperative attack | _Simulation |
| [13] | Hybrid | Anomaly-based | DoS | _Simulation |
| [9] | Distributed | Signature-based | Cyber-security attack | _None |
| [11] | Hybrid | Hybrid | Cyber-attack | _Simulation |
| [14] | Hybrid | Hybrid | Vital attacks | _Simulation |
| [15] | Distributed | Signature-based | MIMA, replay and impersonation attack | _Simulation |
| [16] | – | Signature-based | DoS attacks | _Empirical |
| [17] | Hybrid | Anomaly-based | DDoS | _Empirical |
| [18] | – | Signature-based | DoS attacks | _None |
| [19] | Distributed | Signature-based | Distributed attack | _Simulation |

# 3 HIS

HIS is the basic protection system of human that supports the human to survive environmental threats and diseases. Moreover, by resembling the internet to human beings in a number of ways, it can be said that an immune system could be developed for the internet based on the fundamentals of HIS. Immunity system refers to all bodily mechanisms responsible in protecting the body against detrimental agents in the environment such as microorganisms and their products, chemicals, drugs and pollen grains. The HIS is comprised of three defensive lines functioning in cooperation. The first layer consists of skin, mucous membranes, and secretions of skin. As for the second layer, phagocytic white blood cells, antimicrobial proteins, and the inflammatory responses are the subsections. Finally, the third layer, which is the specific defensive mechanisms, includes lymphocytes and antibodies. Antibodies respond to particular microorganisms, aberrant body cells, toxins and other substances marked by foreign molecules, in specific ways. The human immunity system is of two types: Innate immunity system, and acquired immunity system [20, 21]. Figure 3 illustrates the lymphoid organs, and their main functions.

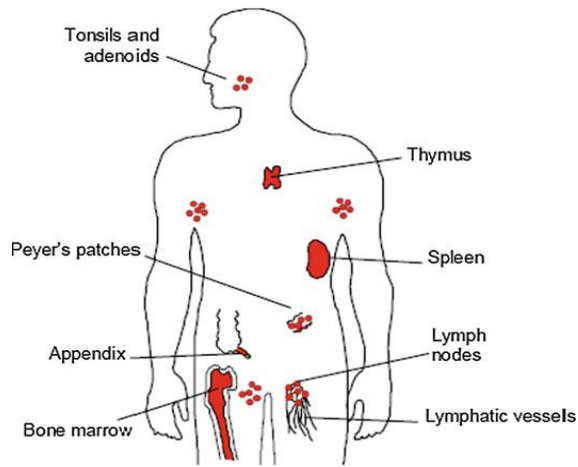

Figure 3: HIS structures [22]

## 3.1 HIS Algorithm

As discussed in the previous section, the HIS is a rather complicated mechanism, capable of protecting the body against a tremendous set of unnecessary pathogens. This mechanism HIS is constructed upon is exceptionally efficient, most notably in self and non-self-antigens distinction. A non-self-antigen is any external factor that can initiate an immune response such as a bacteria or an attack. On the other hand, self-antigens are on the opposite side of non-self-antigens. Self-antigens are the living organism's own cells. The five main theories regarding HIS algorithms are affinity, clonal selection, and negative selection.

- *Affinity*: Different models are used in HIS to calculate the affinity between antibodies (defending) and antigens (attacking). The affinity model is very important because the capability of detection relies on the affinity between the antigen and the detector. Let's assume that the coordinates of an antibody are given by $Abi = (Abi1, Abi2, ..., Abin)$ and the coordinates of an antigen are $Agi = (Agi1, Agi2, ..., Agin)$; the distance between them, $D$, is the affinity [23]. We can use Eq. (1) to calculate $D$.

$$\text{Let } R = \{r_0, r_1, ..., r_m\}, \quad S = \{s_0, s_1, ..., s_m\} \tag{1}$$

$$\text{Manhattan } D = \sum_{i=0}^{m} |r_0 - s_0|$$

$$\text{Hamming } D = \sum_{i=0}^{m} \delta = 1 \text{ if } r_0 {}^1 s_0, \text{ if otherwise}$$

$$\text{Euclidean } D = \sqrt{\sum_{i=o}^{m} (r_0 - s_0)^2}$$

where *r* and *s* respectively represent different characteristics of columns, *m* is the number of data.

- *Clonal Selection Algorithm*: Clonal selection describes a response to an antigen by the immune system. The antibodies that can recognize the antigens multiply and are chosen over ones that do not. This allows detectors to clone their parents by a mutation mechanism with high rates while the antibodies which are self-reactive get eliminated. This act is known as clonal selection. CLONALG was the algorithm created by De Castro which is based on clonal selection [24]. This algorithm takes into account all counts about cloning the best antibodies, affinity maturation, taking out non-stimulated antibodies, and maintaining diversity. Clonal selection has a great strategy for optimization and pattern recognition. This helps evolve the immune system, so it can recognize the antigens that it met in the past.
- *Negative Selection Algorithm*: According to a general definition, a negative selection algorithm is any algorithm with classification aim that imitates the negative selection process in the vertebrate immune system; an idea proposed and developed first by Forrest et al [25]. Such algorithms are part of one-class classifiers; meaning their training sessions are performed on un-labelled data, which are sampled from a certain sub-region of the problem domain. Then, the results are utilised to determine whether the new and unseen data points belong to the same sub-region or not. The algorithms for negative selection are based on agents that are called detectors, which can patterns matching with small problem domain subsets. For instance, in the problem domain illustrated as $U = \{0,1\}^5$, consisting binary strings of length 5, the pattern $1***1$ is a candidate for being a detector, since it matches with all the strings starting and ending with 1. [23].
- *Learning:* In an HIS, agents learn to distinguish, through actions such as negative selection, clonal selection, danger theory, or human immune networks, self and non-self [24].
- *Adaptation:* The immune system is capable of initiating attacks on invaders that the innate defensive system is incapable of removing, via an acquired or adaptive immunity system. Such adaptive immune system can be directed to specific types of invasions, and can be modified via being exposed to agents imposing such invasions. It generally includes white blood cells named lymphocytes. In specific, B and T-cell lymphocytes facilitate the recognition process and aid in destroying specific types of substances. Immunogens (or antigens) are substances that can generate response from the lymphocytes. However, antigens do not attack the microorganisms themselves. Instead, substances existing in the microorganisms including toxins or enzymes are considered foreign by the immune system. Responses created by the immune system that are of adaptive nature normally target the provoking antigens and thus, are referred to as antigen-specific [24].

In Algorithm 1, the pseudo code for Negative Selection Algorithm is demonstrated [26].

```
Algorithm 1: Pseudo code for Negative Selection Algorithm
 1: Procedure Negative Selection Algorithm
 2: Input: A S ⊂ Ui ("self-set"); a set Mo ⊂ Ui ("monitor set"); an integer ni
 3: Output: For each element mo ∈ Mo, either "normal UAV" or "malicious UAV".
 4:     // phase I: Training
 5:       de ← empty set
 6:       while |De| < ni do
 7:              de ← the random detector set
 8:              if de does not match any element of Si then
 9:                  insert de into De
10:              End if
11:       End while
12:     // phase II: Classification
13:           For each mo ∈ mo do
14:              if mo matches any detector de ∈ De then
15:                  output "mo is non-self" (an attacker)
16:              else
17:                  output "mo is self"
18:              End if
19:           End For
20s: End Procedure
```

## 4 The proposed SUAS-HIS framework

In the following section, we design a cyber-security threats-immune schema by employing the HIS algorithm. The proposed system consists of five steps, such as SUAS-HIS network model is discussed in Sect, 4.1. Next section defines the information exchange pattern of UAS. Section 4.3 describes Motion direction of UAV. Details of SUAS-HIS schema is discussed in Sect. 4.4. Security and performance evaluation is discussed in Sect. 4.5, and major parts of the immune system and conducted mapping is discussed in Sect. 4.6.

### 4.1 SUAS-HIS network model

We consider a UAS network, in which UAVs are deployed in an infinite three-dimensional (3D) Euclidean space according to a homogeneous Poisson point process (PPP). The UAVs have a maximum one-hop communication range. A UAV can transmit information to the intended destination UAV directly, or via a relay by one or more UAVs. The multi-hop scheme is decode-and-forward, in which the relaying UAV decodes an arriving packet and then transmits to the next hop. Also, in SUAS-HIS network model, a safe solution is provided to secure the UAVs, which is effective on two ways: First, it has high detection accuracy and low false positive and negative rates and second, it detects and isolates attacks quickly. In the proposed method, the security issues such as WH, BH, GH and FID attacks that could target the UAV are prevented. In order to detect the attacks with a high accuracy, other features could be added in addition to those of Table 3.

## 4.2 Information exchange pattern of UAS

A typical process of exchanging information is as follows. Initially, a source ground station ($G_{src}$) delivers a message to a UAV ($U_1$). This UAV then travels a distance ($D_1$), in order to meet and deliver the message to another UAV ($U_2$). Next, this UAV delivers the message to another UAV ($U_3$), and it continues in the same manner until the final UAV ($U_N$) delivers the message to the ground station in the destination ($G_{DST}$). With the aim of minimizing the latency in delivering end-to-end packets, each UAV participating in this process is mandated to meet the next UAV precisely at the designated time.

In real environments however, since the UAVs have different performances due to variations in engines and environmental uncertainties, they fly at different velocities and therefore, it is unrealistic to expect that all UAVs will be capable of following a similar pattern. For instance, UAVs with higher speeds may travel longer distances compared to others. In addition, the communication line between two UAVs can only be established when they are in the communication range. Therefore, it is mandatory to develop a collaboration among UAVs. Once two UAVs are in the communication range, or they are in the same connection area, they can exchange data packets. This process is time consuming. When a UAV is travelling outside the connection area, or in a connectionless area, no data packet exchange is likely to occur.

## 4.3 Motion Direction of the UAV

To provide motion for the UAVs, this study employed the smooth turn (ST) mobility model. Using ST enables the UAVs to have smoother trajectories, which includes flying in straight trajectories or making turns with larger radius. Therefore, ST is being widely employed for the purpose of analyzing UASs. This model is capable of capturing the acceleration correlation of the UAVs in both temporal and spatial domains, and is accommodating for the purpose of design and analysis. As stated in [27], there exists a uniform distribution for the stationary node of the ST model, which results in a series of closed-form connectivity.

Table 3: Cyber security threats features

| Cyber security threats | Features |
| --- | --- |
| Wormhole attack | Data injection rate |
| Black hole attack | Data injection rate |
| Gray hole attck | Data injection rate |
| FID | Messages modified rate |

The major acronyms and notations used in this paper are provided Table 4.

Table 4: Major acronyms and notations used in this paper.

| Acronyms | Abbreviated acronyms | Notation | Abbreviated notations |
|---|---|---|---|
| $NS-3$ | Network Simulator 3 | $RREQ$ | Route Request |
| $NAM$ | Network Animator | $RREP$ | Route Reply |
| $IDS$ | Intrusion Detection System | $DA$ | Denoising Auto Encoder |
| $HIS$ | Human Immune System | $UAV_S$ | Source $UAV$ |
| $SUAS-HIS$ | Securing of Unmanned Aerial Systems by Human Immune System | $UAV_D$ | Destination $UAV$ |
| $UAS$ | Unmanned Aerial Systems | $T_s$ | Surveillance Threshold |
| $UAV$ | Unmanned Aerial Vehicles | $DR$ | Detection rate |
| $AVN$ | Aerial Vehicles Networks | $G_{src}$ | Ground station source |
| $U2U$ | UAV-UAV | $G_{DST}$ | Ground station destination |
| $U2S$ | UAV-Satellite | $m$ | Malicious |
| $U2T$ | UAV- Traffic control tower | $P_m(r)$ | Probability malicious (route) |
| $U2G$ | UAV-Ground station | $SSI$ | Signal Strength Intensity |
| $GPS$ | Global Positioning System | $F_r$ | Fitness route |
| $WH$ | Wormhole | $P_{sb}$ | Probability subversive behavior |
| $BH$ | Blackhole | $N_s$ | Number of packets |
| $GH$ | Grayhole | $RTT_i$ | Round Trip Time |
| $FID$ | Fake Information Dissemination | $MaxRTT$ | Maximum RTT |
| $FPR$ | False positive rate | $SSI_i$ | Signal Strength Intensity i |
| $FNR$ | False negative rate | $MaxSSI$ | Maximum SSI |
| $TPR$ | True positive rate | $Agi$ | Anti-gen |
| $TNR$ | True negative rate | $Abi$ | Anti-body |
| $PC$ | Packet counter | $D$ | Distance |
| $SFA$ | Security Framework Aircraft | $R$ | Route |
| $ROS$ | Robot Operating System | $AS$ | Antigen Self |
| $DoS$ | Denial of Service | $SPC_{UAV}$ | The number of packets sent |
| $ACO$ | Ant Colony Optimization | | |
| $RTT$ | Round Trip Time | | |

## 4.4 Details of the proposed method: Securing the UASs against security threats using HIS

After the source UAV or ground station receives the first response message, the default air systems detect the shortest route and sends packets by receiving the first message from other UAVs. Accordingly, it pays no attention to the security of the communication channel (the absence of a WH, BH, a GH and FID attacks). However, in the proposed method, the UAV examines the security of routes by the HIS algorithm and then sends packets through the safest route. The overall idea of the proposed method is that as in the HIS, $Abi$ are taught to discover and remove malignant $Agi$, here a set of rules are created and updated to track the routes infected by attacking UAVs and exclude them from the cycle. The proposed mapping and comparison between the HIS and UAS under the attack is shown in Table 4. The details on how to use the mapping are provided in Table 4.

Table 4: The mapping between HIS and UASs.

| Human Body | UASs |
|---|---|
| Antigen Structure | The set of all routes connecting the source to destination UAV. |
| Antibody Structure | Attack-detection rules based on (SSI, RTT), and packets count backwards from the destination). |
| Colonization | Reproduction of antibodies that best match the antigens. |
| Affinity value | The sum of SSI, RTT, and packets count backwards from the destination. |
| Mutation | Under identical conditions, the fastest route response is selected. |

## 4.5 Assessing the safety and performance of the routes with UAVs

At this stage, the intended antibodies are designed to provide security and performance for both UAVs and ground stations simultaneously. This design is done in the following three phases:

**Phase 1: Initial assessment of candidate routes using Hello packets**

At this phase, each route for which the route response is received is tested in terms of security. Accordingly, a "Hello packet" is sent through each route, and the target UAV is required to send a confirmation packet for those routes containing the UAVs if the "Hello packet" is received. Obviously, if a route is affected with malicious UAVs, the "Hello packet" will not be delivered and the confirmation packet will not be received. Under these conditions, the probability of rout contamination increases i.e. the $P_m(r)$ value for the r route increases but if the "Hello packet" reaches its destination, a confirmation packet will be received, which means that there is no malicious UAVs, so the $P_m(r)$ value decreases. The process of sending "Hello packet" is repeated 4 times.

**Initial value of the $P_m(r)$ variable:** If, according to the attacker UAV detection mechanism the route is confirmed, the initial $P_m(r)$ value will be zero but if the route is not approved, $P_m(r)$ value will be 100. Then, in order to update it, the source UAV sends a "Hello packet" four times from all possible routes to the destination UAV. If the confirmation packet is sent by the destination UAV, 25 units will be deducted from the $P_m(r)$ variable; but if the confirmation packet is not sent by the destination UAV, 15 units will be added to the $P_m(r)$ variable. This is repeated 4 times and $P_m(r)$ is updated for all routes (threshold value). If the $P_m(r)$ value is greater than 50, that route will be set aside as a contaminated route (rejected). Otherwise, that route will be sent to the next part i.e. Phase 2 to evaluate the safety and efficiency of the routes.

**Phase 2: Assessing and discovering routes containing malicious UAVs**

**Reverse packet counting from destination UAV:** As the data packet reaches the destination UAV, the destination UAV sends a query response packet to the UAV at the source route (data transmission route) at a distance of 2-hops. If the following route indicates the route between the source and destination UAVs, $UAV_S.UAV_0.UAV_1.UAV_2.UAV_{(n-3)}.UAV_{(n-2)}.UAV_{(n-1)}.UAV_n.UAV_D$, then the $UAV_D$ sends a packet to the $UAV_{(n-1)}$ located 2 hops away from the $UAV_D$. Query request is used to find the number of data packets sent by a UAV to the UAV used in its next hop. $UAV_{(n-1)}$ sends a query packet to the destination $UAV_D$. The response to query packet, which contains the number of data packets sent by a UAV on the source route, is sent to the UAV by its adjacent hop. The destination UAV checks the query response to see whether the previous hop (for example $UAV_n$) has sent all data packets received from its previous $UAV_{(n-1)}$. If not correct, the destination UAV will add both $UAV_{(n-1)}$ and $UAV_n$ to the suspect list. If it is correct, it means that the two UAVs participated in sending data correctly. Therefore, the destination UAV re-sends a new query request to the $UAV_{(n-3)}$ in the 2-hop distance from the $UAV_{(n-1)}$ on the source route. Based on the received query request, the destination UAV explores whether $UAV_{(n-3)}$ and $UAV_{(n-2)}$ have sent all data packets they have received or not. This process continues until the query request reaches the UAV that the UAV has no previous hop at 2-hops away on the source route. Using the query response packets, the first row examines the behavior of sending the intermediate UAVs' data on the route with high reliability from the source UAV through the destination UAV. If the difference in the number of packets transmitted between both intermediate UAVs exceeds the surveillance threshold, the destination UAV will mark both intermediate nodes as suspicious UAVs. The probability of subversive behavior $P_{sb}$ between each of the two UAVs is calculated in accordance with Eq. (2).

$$P_{sb} = \left( \frac{SPC_{UAV_{n-2}UAV_{n-1}} - SPC_{UAV_{n-1}UAV_{n-2}}}{N_s} \right) \quad (2)$$

Where, $SPC_{UAV_{n-2}UAV_{n-1}}$ represents the number of data packets transmitted by $UAV_{n-2}UAV_{n-1}$. If $P_{sb} > T_s$, both of them $UAV_{n-2}$ and $UAV_{n-1}$ are malicious. Here the surveillance threshold is $T_s$ and could range between $0$ and $0.2$, and $N_s$ represents the packets sent from the source $UAV$. The $T_s$ is for monitoring the packet loss rate in UAS networks. This means with how many lost packets in the network, we seek to discover malicious UAVs. If $T_s$ is set to zero, it means that we expect all packets to reach the destination successfully and no packet to be lost. However, if $T_s$ is set to 0.1, it means that it does not matter if up to 10 percent of the packets get lost and no action will be taken for discovering malicious UAVs but if it is more than 10 percent, we must seek to discover malicious UAVs according to the proposed method. Likewise, considering the value to be 0.2 means 20 percent of all the transmitted packets. Setting the surveillance threshold depends on the network size and the number of transmitted packets.

**Phase 3: Assessing the safety and efficiency of the routes that went through Phase 1 and Phase 2**

Given that, the attacks show the number of hops below the actual value, if a route has a lower number of hops, the probability that the route is infected increases. Meanwhile, the low $RTT$ and high $SSI$ of the UAVs on the route, in addition to improve the route safety in terms of attacks, it is also desirable based on the route performance. Therefore, the optimal route can be considered as a route that maximizes the $F_r$ index according to Eq. (3).

$$F_r(RTT_i, SSI_i) = \left( \frac{MaxRTT}{RTT_i} \right) + \left( \frac{SSI_i}{MaxSSI} \right) \quad (3)$$

The meaning of the equation $\left( \frac{MaxRTT}{RTT_i} \right)$ is that the routes which have a lower $RTT$ will be chosen as the intended route.

Since among the values equation $\left(\dfrac{MaxRTT}{RTT_i}\right)$ yields as output for all the routes, the maximum value is selected as the shortest route.

> ***Example:*** $RTT(route\ 1) = 5;\quad RTT(route\ 2) = 10;\quad RTT(route\ 3) = 20$
>
> According to the equation $\left(\dfrac{MaxRTT}{RTT_i}\right)$, the output will be as follows:
>
> $Max\ RTT(all\ route) = 20;\quad 20/5 = 4;\quad 20/10 = 2;\quad 20/20 = 1$
>
> As a result, according to the equation, the bigger number means that it had a smaller *RTT* and it will be selected as the output.
>
> The meaning of the equation $\left(\dfrac{SSI_i}{MaxSSI}\right)$ is that the routes with higher *SSI* get selected as the intended route. Since among the values this equation yields for all routes, the maximum value is selected as the higher *SSI*.

> ***Example:*** $SSI(route\ 1) = 5;\quad SSI(route\ 2) = 10;\quad SSI(route\ 3) = 20$
>
> According to the equation $\left(\dfrac{SSI_i}{MaxSSI}\right)$, the output will be as follows:
>
> $Max\ SSI(all\ route) = 20;\quad 5/20 = 0.25;\quad 10/20 = 0.5;\quad 20/20 = 1$
>
> Therefore, according to the equation, a bigger number means it had a bigger *SSI* and gets selected as the output.
>
> By combining the *RTT* and *SSI* equation, the equation $F_r(RTT_i, SSI_i) = \left(\dfrac{MaxRTT}{RTT_i}\right) + \left(\dfrac{SSI_i}{MaxSSI}\right)$ was obtained. For the output of this equation, the maximum number will be the best and the most secure route.

$$UAV_{SR} = (1 - P_m(r)) * F_r(RTT_i, SSI_i) \tag{4}$$

In Eq. (4), the $P_m(r)$ variable calculates the probability of the $r$ candidate routes being malicious. On the other hand, $(1 - P_m(r))$ presents the probability of the $r$ candidate routes being healthy. Also, equation $F_r(RTT_i, SSI_i)$ selects the route which has a low *RTT* and high *SSI*. By combining these two, we created the $UAV_{SR}$ equation in order for routes to be selected from the $r$ candidate ones which are less likely to be malicious. Therefore, the chosen route is the one which has both security and high performance.

Since malicious UAVs always insert their location incorrectly in the RREP packet, due to this problem, we use RTT to obtain the precise distance between UAVs in order to forward the packet to the UAV which has the shortest distance. Also, by calculating SSI for each UAV, we notice their fake signals. In order to deliver the desired package to a healthy UAV. By doing so, malicious activities of WH, BH, GH, and FID attacks will be prevented.

> *Example:*
> $P_m(route\ 1) = 0.25;\quad P_m(route\ 2) = 0.50;\quad P_m(route\ 3) = 0.40;$
>
> According to the $UAV_{SR} = (1 - P_m(r)) * F_r(RTT_i, SSI_i)$ equation, it will be as follows:
>
> $Route1 = (1 - 0.25) * 5 = 3.75$
>
> $Route2 = (1 - 0.50) * 2.5 = 1.25$
>
> $Route3 = (1 - 0.25) * 1.25 = 0.9375$
>
> As seen in the examples, the route which had a lower *RTT* and a higher *SSI* was selected as the output.

## 4.6 Major parts of the immune system and conducted mapping

The proposed method is structured according to the flowchart shown in Figure 4: random population generation of antibodies (rules), Antigen structure (set of all routes from the target UAV to the destination UAV), matching, rejection of matched unsafe communication channels, completion of detector set (rules that detect routes infected by the attacker), storing in the safety memory and detector hyper mutation.

**Antibody and antigen structure:** Antibody is considered as the three features of limiting the attacks (RTT, SSI and reverse packet counting from destination) and antigen is the total set of routes for which the route response are achieved.

**Affinity:** As defined in the HIS, antibodies that have a stronger binding to an antigen than other antibodies are selected as a safe antibody for composition. In the proposed method, affinity is selected as routes with less RTT, high SSI and reverse packet counting from destination. Otherwise, the candidate route will be eliminated.

**Match:** In this section, the responses of the routes collected on the UAV of the source are compares in terms of the two below features and the safest route is selected. Meanwhile, the most important feature of the attacker detection mechanism is that they are modified over time and defined in a way that can be easily corrected and learned.

*Details of the first and second features of the antibody in matching section:* These details are expressed in two following features:

*First feature (RTT between the source UAV and destination UAV):* The RTT is calculated for all routes received from the source UAV to destination UAV. The objective of the first feature is to calculate the precise and correct distance between UAVs in order to prevent the malicious operation of the intruder UAV which makes its distance seem less. This is done using RTT.

*Second feature (SSI):* The attacker UAV generates high-intensity signal to control the target UAV. In such detection, it collects all SSIs generated by the transmitters UAVs and then compares them with normal UAV SSIs. Accordingly, the suspicious transmitter signal and normal UAV signals are differentiated. The objective of the second feature is to distinguish the SSIs created by the malicious UAV from the ones created by healthy UAVs. This is done this way that we calculate the difference between the SSI with the highest signal and all other received SSIs. Now, if this difference is bigger than the threshold value, we conclude that the considered SSI is not safe and remove it. this step is performed on all received SSIs. The reason for comparing to the threshold value is that the intruder UAV wants to make the destructive attack (4 attacks we mentioned before) as close to the source UAVs as possible. This is because the malicious UAV being closer is the most dangerous type of attack since it will remove the most amount of data traffic.

**Completion of detector set:** For all routes the $P_m(r)$ value of which is less than 50, the algorithm 2 is used to choose the safest route. The pseudo code of the SUAS-HIS is given in Algorithm 2.

According to algorithm 2, the $F_r$ function is calculated for each route and the route that has a larger number is selected as the safest route.

**Hyper mutation:** Among the assessed routes, routes with almost identical conditions are transmitted to the hyper mutation stage so that the routes are assessed with a different criterion (under the equal terms, the fastest route response is selected) to select the safest route for UAVs.

**Registration in safety memory:** routes that have the largest number in accordance with Eq. 3, are the safest route and are registered in the memory for later use.

---

**Algorithm 2:** Pseudo code for SUAS-HIS schema

1: **Initialize** the Antigen collection time to 10s;
2: **Initialize** the Antigen towards min to 70s;
3: **Initialize** the Delay buffer size max to 1200;
4: **Initialize** the Storing time to 11s;
5: **Initialize** the Max number of antigens to 1200;
6: **Let** $P_m(r)$ the probability of the infected route;
7: **Let** $R$ denote the number of candidate route;
8: **Let** $F_r$ denote the fitness route;
9: **Let** $UAV_{SR}$ denote the immune route;
10: **Procedure** Selecting immune route
11:     For $r = 1$ To $R$ Do
12:         IF $P_m(r) > 0.5$ || $Attack = Confirmed$ Then
13:             Push out $route_r$
14:             Broadcast $route_r$ to Inform other UAVs
15:         Else
16:             $F_r(RTT_i, SSI_i) = \left(\frac{MaxRTT}{RTT_i}\right) + \left(\frac{SSI_i}{MaxSSI}\right)$
17:             $UAV_{SR} = (1 - P_m(r)) * F_r(RTT_i, SSI_i)$
18:             IF $UAV_{SR} < 0.5$ || $Attack = NotConfirmed$ Then
19:                 Broadcast $route_r$ to Inform other UAVs
20:             EndIf
21:         EndIf
22:     EndFor
23: **End Procedure**

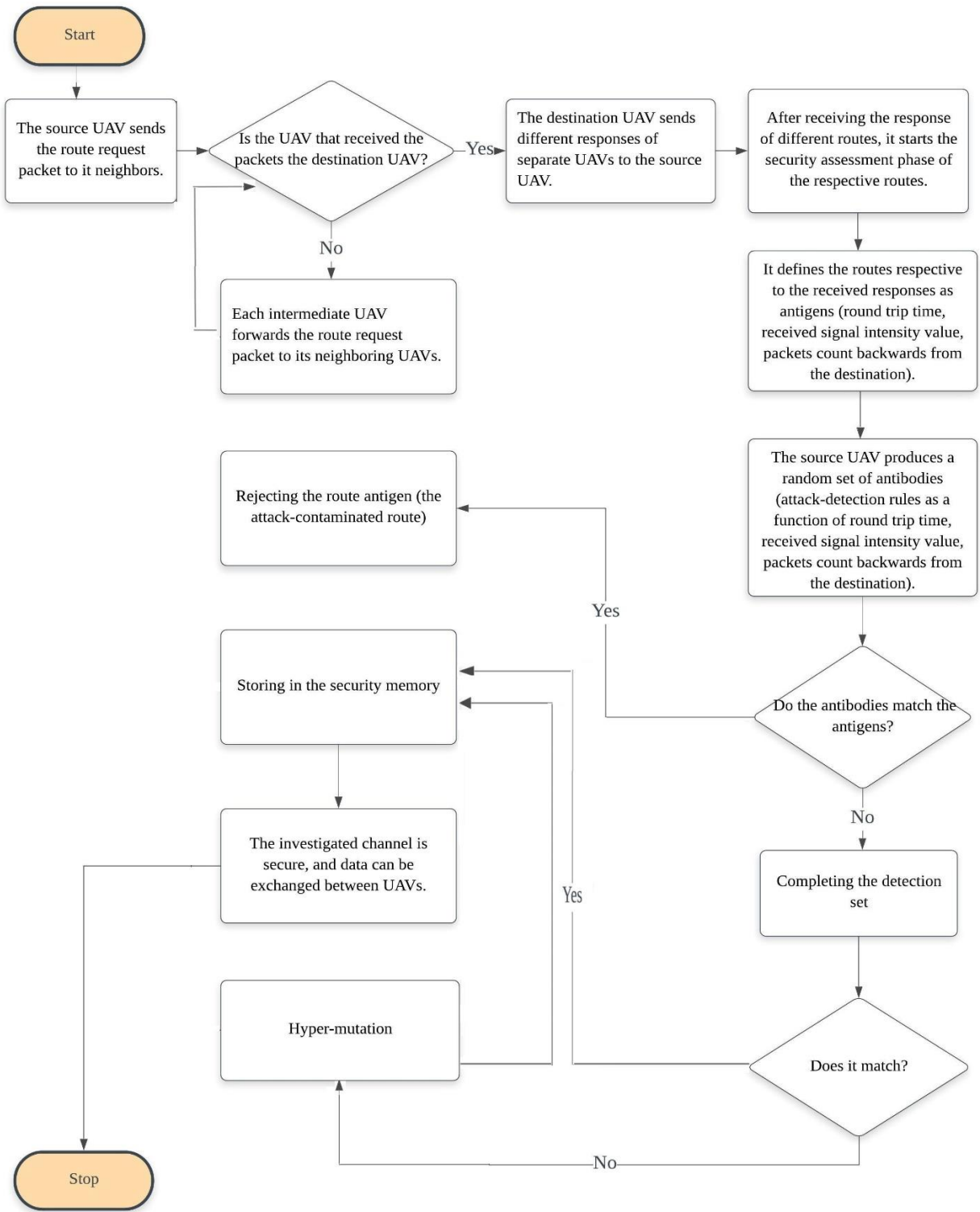

Figure 4: Flowchart of the SUAS-HIS

## 5  Evaluating the Performance

To prevent threats of cyber security, we evaluate our proposed SUAS-HIS approach's performance in the following section.

## 5.1 Performance metrics

In this section, the effectiveness and performance of our proposed SUAS-HIS approach is thoroughly evaluated with comprehensive simulations. The results are compared with Cyber Security System, BRUIDS, and SFA approaches proposed in [9], [11] and [10], respectively. The false positive, false negative and detection ratio are evaluated.

### 5.1.1 False positive rate

FPR is defined as the division of the number of misbehaving UAVs being misdirected, to the overall count of UAVs currently well-behaving, as illustrated in Eq. (5).

$$FPR = \left(\frac{FPR}{FPR+TNR}\right)*100 \quad \text{Where:} \quad TNR = \left(\frac{TNR}{TNR+FPR}\right)*100 \tag{5}$$

### 5.1.2 False negative rate

FNR is the number of UAVs that are currently well-behaving, but are considered as malicious by the trust system, as demonstrated by Eq. (6).

$$FNR = \left(\frac{TPR+TNR}{All}\right)*100 \quad \text{Where:} \quad TPR = \left(\frac{TPR}{TPR+FNR}\right)*100 \tag{6}$$

### 5.1.3 Detection rate

DR is defined as the division of the number of detected UAV currently misbehaving, to the overall number of actual nodes currently misbehaving. In other words, it is defined as the probability of successful identification of all security threats, as demonstrated in Eq. (7).

$$DR = \left(\frac{TPR}{TPR+FNR}\right)*100 \quad \text{where} \quad All = TPR+TNR+FPR+FNR \tag{7}$$

### 5.1.4 Packet Delivery Ratio

As defined, Pocket Delivery Ratio results from dividing the total received packets of data at the destination UAV, to the total transmitted packets of data by the source UAV, denoted in percentage [28-30]. Eq. (8) demonstrates the average obtained PDR for $N$ experiments.

$$PDR = \frac{1}{n} * \frac{\sum_{i=1}^{n} X_i}{\sum_{i=1}^{n} Y_i} * 100\% \tag{8}$$

## 5.2 Simulation setup and comparing algorithms

The difficulties in implementation and debugging UASs in real networks, raises the necessity to consider simulations as a fundamental design tool. The main advantage of simulation is simplifying analysis and protocol verification, mainly in large-scale systems. It is possible to employ a NAM in the Network Simulator (NS-3) [31, 32] to visualise the results. In this section, the performance of our proposed approach is evaluated using NS-3 as the simulation tool, and the results are discussed further. It is worth mentioning that all SUAS-HIS, Cyber Security System, BRUIDS and SFA parameters and settings are considered to be equal [33]and [34, 35].

## 5.3 Simulation results and Analysis

In the following section, the performance of SUAS-HIS in terms of security is analyzed presented in Table 5 [36]. In this paper, simulation with two different scenarios is evaluated for UAS networks. The first scenario has a variable number of UAVs from 100 to 400, fixed simulation time of 1400 seconds, and a network size of 6000 by 6000. The second scenario has a fixed number of 400 UAVs, variable simulation time from 200 to 1400 seconds, and a network size of 4000 by 4000. Each scenario is run 4 times (according to a, b, c, and d diagrams). The first time, it is evaluated with a %7 malicious UAV rate, the second time with a %14 malicious UAV rate, the third time with a %21 malicious UAV rate, and the fourth time only with a variable simulation time from 200 to 1400. The UAV flies with constant speed of 180 m/s. In addition, some significant parameters employed are enlisted in Table 5.

Table 5: Setting of simulation parameters.

| Parameters | Value |
| --- | --- |
| First scenario | Number of UAVs: 100 to 400 |
| | Simulation time: 1400 S (fixed) |
| | Network size (m x m): 6000 x 6000 |
| Second scenario | Number of UAVs: 400 (fixed) |
| | Simulation time: 200 to 1400 S |
| | Network size (m x m): 4000 x 4000 |
| Channel type | Channel/Wireless channel |
| MAC Layer | MAC/802.11.b |
| Traffic type | CBR |
| UAV speed | ١٨٠ m/s |
| Transmission layer | UDP |
| Packet size | 512 Byte |
| Malicious UAV | 0 - 0.30 |
| Type of attacks | FID, BH, GH, and WH |
| Transmission range | 30 M |

Table 6-9 compares the performance of SUAS-HIS with that of Cyber Security System, BRUIDS and SFA in terms of FPR, FNR, DR, and PDR.

Table 6: *DR* (in %) of various frameworks with varying degree of malicious UAVs.

| Misbehaving UAV ratio | DR (%) | | | |
| --- | --- | --- | --- | --- |
| | *Cyber Security System* | *BRUIDS* | *SFA* | *SUAS – HIS* |
| 0 | 96.63 | 95.3 | 95.2 | 99.5 |
| 0.05 | 94.49 | 94.5 | 93.57 | 98.2 |
| 0.10 | 85.46 | 86.1 | 86.8 | 96.38 |
| 0.15 | 78.35 | 79.5 | 81.37 | 94.27 |
| 0.20 | 69.19 | 72.8 | 75.43 | 92.28 |
| 0.25 | 55.34 | 62.7 | 71.16 | 89.7 |
| 0.30 | 51.14 | 59.7 | 65.67 | 87.4 |

Table 7: *FNR* (in %) of various frameworks with varying degree of malicious UAVs.

| Misbehaving UAV ratio | FNR (%) | | | |
|---|---|---|---|---|
| | *Cyber Security System* | *BRUIDS* | *SFA* | *SUAS – HIS* |
| 0 | 2.93 | 4.5 | 4.005 | 0.34 |
| 0.05 | 3.43 | 5 | 5.08 | 0.86 |
| 0.10 | 5.19 | 7.1 | 6.3 | 1.27 |
| 0.15 | 10.63 | 9.7 | 8.37 | 2.62 |
| 0.20 | 19.38 | 15.9 | 11.25 | 4.2 |
| 0.25 | 28.2 | 21.4 | 13.76 | 7.83 |
| 0.30 | 34.27 | 24.8 | 19.89 | 9.22 |

Table 8: *FPR* (in %) of various frameworks with varying degree of malicious UAVs.

| Misbehaving UAV ratio | FPR (%) | | | |
|---|---|---|---|---|
| | *Cyber Security System* | *BRUIDS* | *SFA* | *SUAS – HIS* |
| 0 | 3.3 | 3.1 | 4.25 | 0.42 |
| 0.05 | 5.31 | 7.1 | 6.24 | 1.65 |
| 0.10 | 14.05 | 13.9 | 13.01 | 3.41 |
| 0.15 | 21 | 19.7 | 17.35 | 4.57 |
| 0.20 | 29.38 | 26.8 | 24.62 | 7.63 |
| 0.25 | 42.6 | 35.4 | 28.88 | 9.97 |
| 0.30 | 46.67 | 39.4 | 34.09 | 11.83 |

Table 9: *PDR* (in %) of various frameworks with varying degree of malicious UAVs.

| Misbehaving UAV ratio | PDR (%) | | | |
|---|---|---|---|---|
| | *Cyber Security System* | *BRUIDS* | *SFA* | *SUAS – HIS* |
| 0 | 95.9 | 95.3 | 98.1 | 99.7 |
| 0.05 | 82.4 | 85.4 | 91.2 | 95.4 |
| 0.10 | 77.3 | 79.1 | 82.2 | 91.2 |
| 0.15 | 65.3 | 71.3 | 74.3 | 87.3 |
| 0.20 | 60.4 | 62.8 | 67.1 | 82.4 |
| 0.25 | 50.9 | 54.3 | 61.2 | 79.1 |
| 0.30 | 41.6 | 49.3 | 55.3 | 75.4 |

Table 10 represents average values of various frameworks for all metrics under security threats.

Table 10: Average values of various frameworks for all metrics under security threats.

| Schemes | | Detection rate | FNR | FPR | PDR |
|---|---|---|---|---|---|
| Cyber Security System | Number of UAVs (7% of overall nodes) | 77.91 | 10.04571429 | 18.21 | 72.2285714 |
| | Number of UAVs (14% of overall nodes) | 66.98714286 | 13.41714286 | 24.24285714 | 69.6114285 |
| | Number of UAVs (21% of overall nodes) | 50.64857143 | 18.17142857 | 31.24428571 | 64.5428571 |
| | Simulation times | 75.8 | 14.86142857 | 23.18714286 | 0.41 |
| BRUIDS | Number of UAVs (7% of overall nodes) | 80.38571429 | 6.81428571 | 11.7285714 | 73.7857142 |
| | Number of UAVs (14% of overall nodes) | 72.28571429 | 8.38571428 | 18.0571428 | 69.9571428 |
| | Number of UAVs (21% of overall nodes) | 53.24285714 | 14.4571428 | 22.6714285 | 66.8142857 |
| | Simulation times | 74.71428571 | 0.06371428 | 0.046 | 45.7142857 |
| SFA | Number of UAVs (7% of overall nodes) | 89.94714286 | 5.645714286 | 9.332857143 | 77.0285714 |
| | Number of UAVs (14% of overall nodes) | 77.37142857 | 6.605714286 | 14.6 | 71.2857142 |
| | Number of UAVs (21% of overall nodes) | 59.27142857 | 12.71428571 | 18.34 | 68.4914285 |
| | Simulation times | 81.31428571 | 9.807857143 | 18.34857143 | 0.61714285 |
| SUAS – HIS | Number of UAVs (7% of overall nodes) | 98.83 | 0.918571429 | 1.134285714 | 94.4171428 |
| | Number of UAVs (14% of overall nodes) | 92.37 | 3.477142857 | 9.668571429 | 84.4114285 |
| | Number of UAVs (21% of overall nodes) | 86.59285714 | 7.677142857 | 11.15285714 | 77.9128571 |
| | Simulation times | 93.96142857 | 3.762857143 | 5.64 | 0.91428571 |

**FPR:** Figure 5 demonstrates the comparison of the proposed SUAS-HIS framework against three approaches, one game theory-based (SFA), the second adaptive behavior-rule specification-based (BRUIDS), and other one risk-based algorithm (Cyber Security System) frameworks. As shown in the Figure 5(a), when the number of normal UAVs ranged from 100 to 400 and the rate of malicious UAVs increased from 7% to 21%, the FPR generated by the proposed design had a slight and moderate growth compared to the other three designs. The FPR of the proposed design is less than 3% when number of normal UAVs and the rate of malicious UAVs are equal to 400 and 7%, respectively. However, this amount is set to 20% for the SFA, 25% for the BRUIDS, and 35% for the Cyber Security System. The reason for the superiority of the proposed design is the rapid detection of malicious UAVs and their removal by cooperation between ground stations and normal UAVs that the process is performed by the trained rules stored in memory. Also, the other reason is that the SUAS-HIS detects security threats and isolates them from the UAS network, thereby decreasing the FPR that occurs as a result of the attacks. As shown in the Figure 5(a), (b), (c) and (d), SUAS-HIS decreases the FPR by more than 11.98, 14.87, and 25.13% those of SFA, BRUIDS, and Cyber Security System models, respectively.

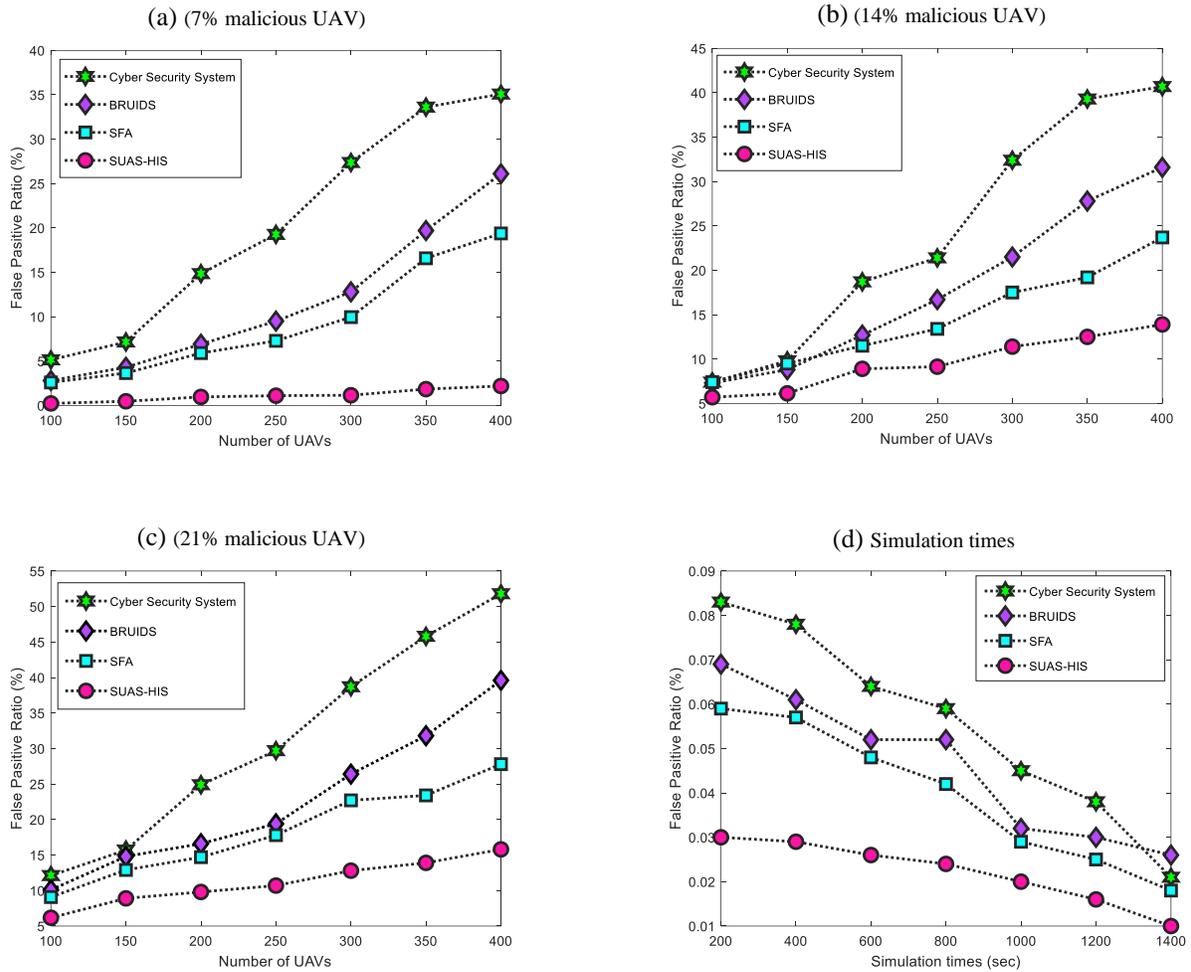

Figure 5: Comparison of the SUAS-HIS proposed scheme, Cyber Security System, BRUIDS and SFA models in term of FPR.

Figure 6 shows the comparison results of the SUAS-HIS proposed scheme, in terms of FPR at different type of attacks and different $T_s$.

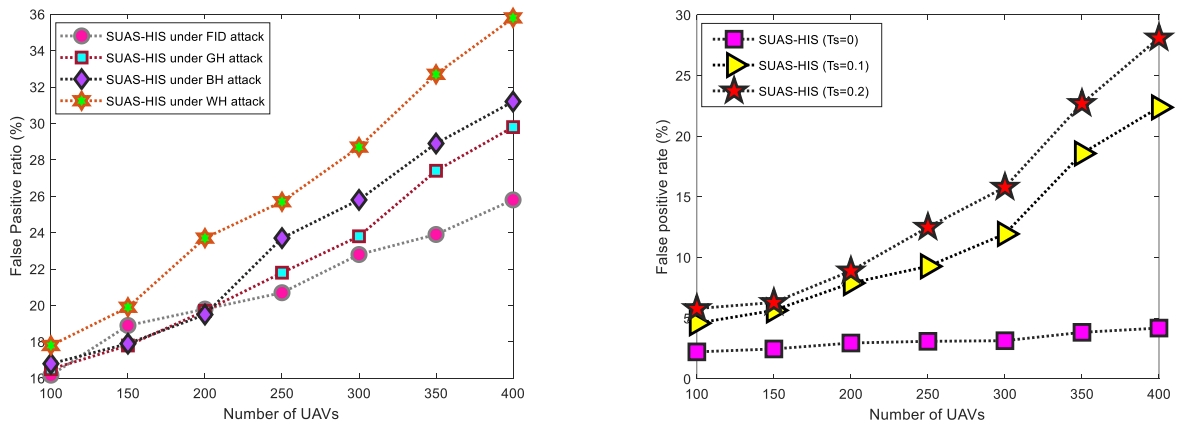

Figure 6: Comparison of the SUAS-HIS proposed scheme, in terms of FPR at different type of attacks and different $T_s$.

**FNR:** Figure 7 shows the comparison of the SUAS-HIS proposed scheme, SFA, BRUIDS, and Cyber Security System models in term of FNR in lethal attacks. (a) Number of UAVs (7% malicious), (b) Number of UAVs (14% malicious), (c) Number of UAVs (21% malicious), (d) Simulation times respectively. As shown in the diagrams, the FNR of the SUAS-HIS proposed schema has increased slightly but this value is much higher in the SFA, BRUIDS, and Cyber Security System. In Figure 7(a), the FNR of the proposed schema is less than 2% when the number of normal UAVs is 400, but for the other three approaches, it is 12, 14 and 18% respectively. In Figure 7(b), when the rate of malicious UAVs is 14%, it is less than 4% in the proposed design but this amount is 7, 12, and 22% for the other three methods respectively. Figure 7(c) explains the FNR under security threats with number of UAVs (21% malicious UAV). The result indicates in Figure 7(d) that, with the traditional technique, the FNR during security threats at simulation time 200 is approximately 6%, which decreases to approximately 4% at 1400 second in a simulation time condition. As shown in the Figure 7(a), (b), (c) and (d), SUAS-HIS decreases the FNR by more than 11.97, 15.3, and 25.74% those of SFA, BRUIDS, and Cyber Security System models, respectively.

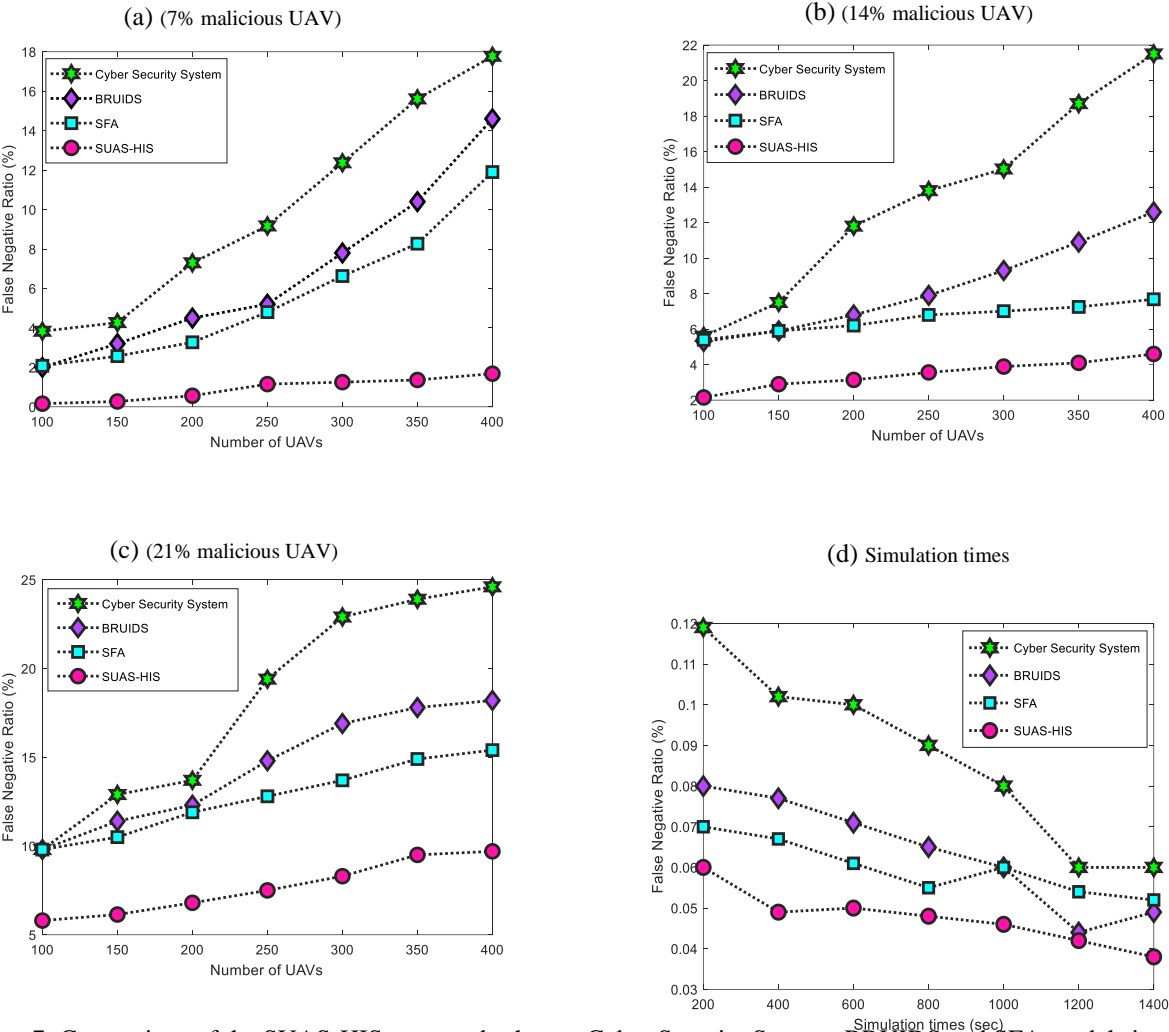

Figure 7: Comparison of the SUAS-HIS proposed scheme, Cyber Security System, BRUIDS and SFA models in term of FNR.

The comparison results of the SUAS-HIS proposed scheme, in terms of FNR at different type of attacks and different $T_s$ are provided in Figure 8.

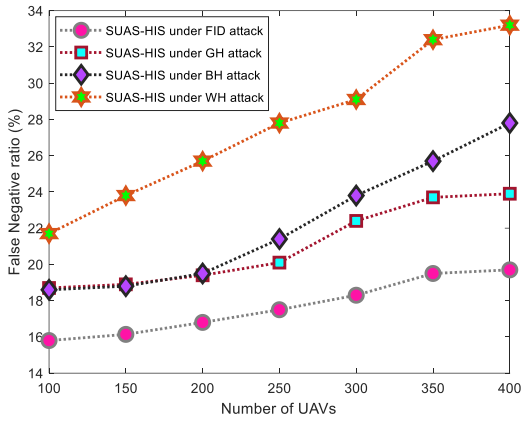
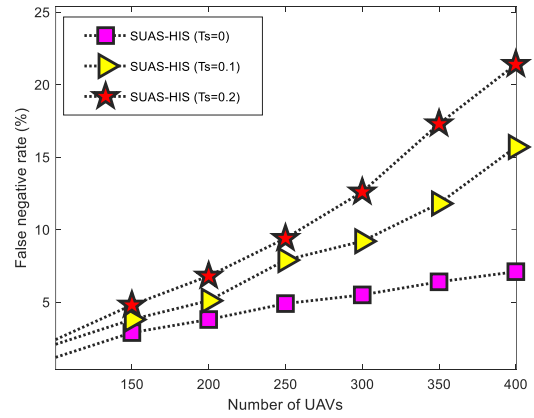

Figure 8: Comparison of the SUAS-HIS proposed scheme, in terms of FNR at different type of attacks and different $T_S$.

**DR:** Figure 9 shows the comparison of the SUAS-HIS proposed scheme, SFA, BRUIDS, and Cyber Security System models in term of DR. (a) Number of UAVs (7% malicious), (b) Number of UAVs (14% malicious), (c) Number of UAVs (21% malicious), and (d) Simulation times respectively. As shown in the Figure (a), the DR in each of the four methods is reduced according to two scenarios, especially when the number of malicious UAVs is high. This reduction is much higher for the Cyber Security System than for other mechanisms. The SUAS-HIS proposed scheme can detect all of the above attacks at a DR higher than 95%. This result is achieved when the number of normal UAVs and the rate of malicious UAVs are equal to 400 and 21%, respectively. The reason for the superiority of the SUAS-HIS is the rapid detection of malicious UAVs and their removal by mapping of unsafe and antigenic routes, which are discovered by the model trained with antibody and removed from the operation cycle. As shown in the Figure 9(a), (b), (c) and (d), SUAS-HIS increases the DR by more than 20.73, 24.4, and 37% those of SFA, BRUIDS, and Cyber Security System models, respectively.

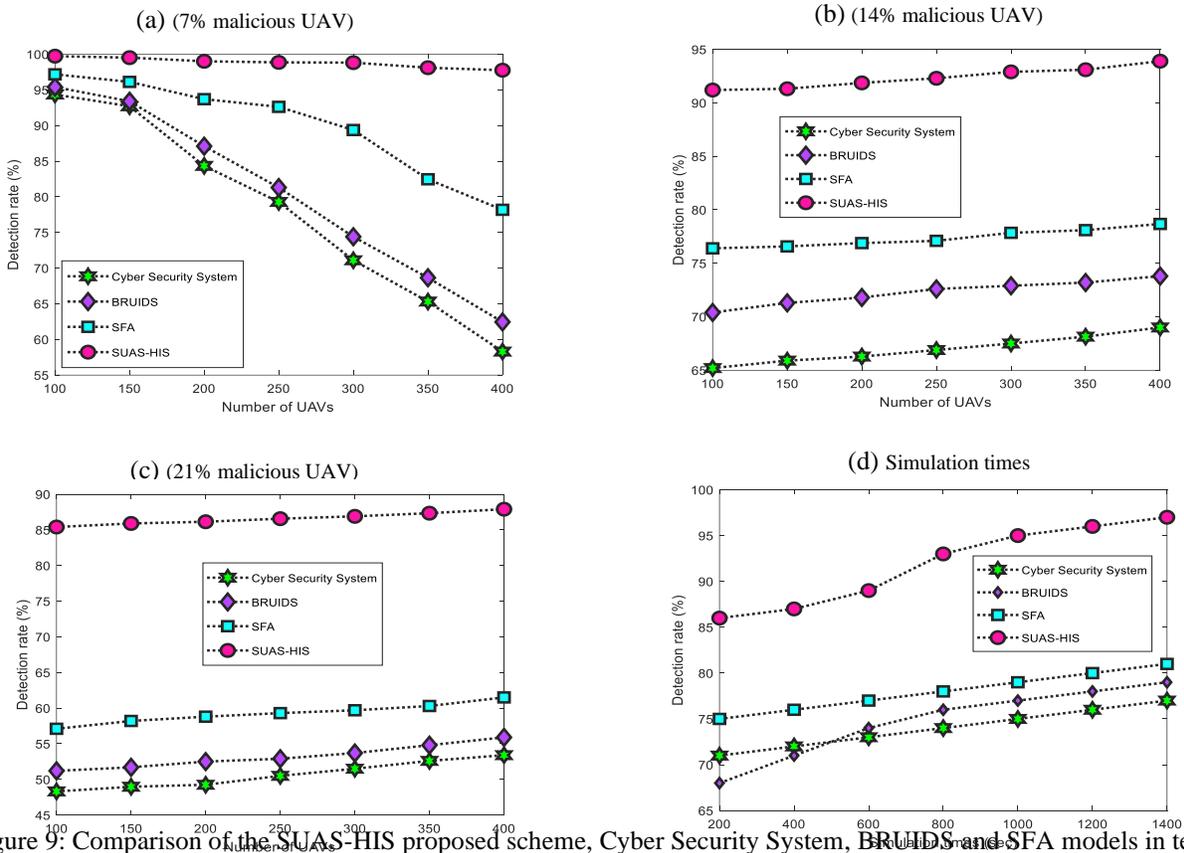

Figure 9: Comparison of the SUAS-HIS proposed scheme, Cyber Security System, BRUIDS, and SFA models in term of DR.

Figure 10 shows the comparison results of the SUAS-HIS proposed scheme, in terms of DR at different type of attacks and different $T_S$.

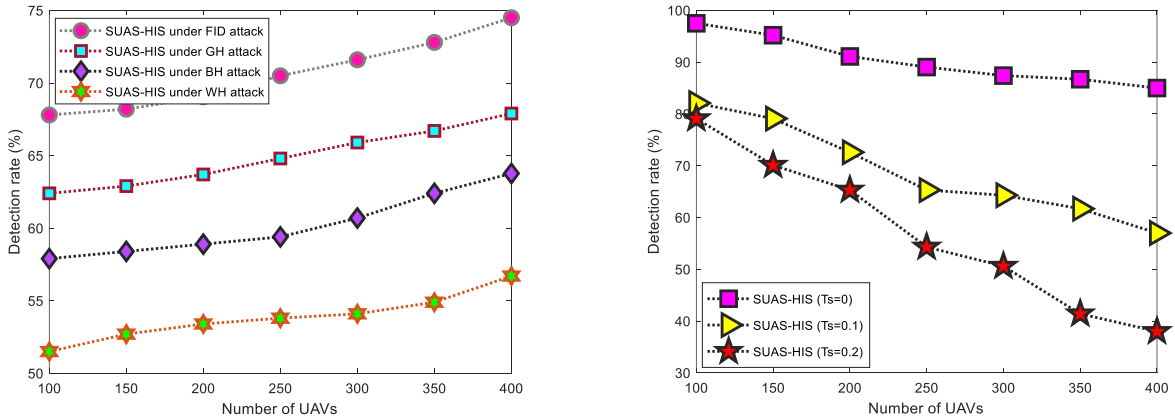

Figure 10: Comparison of the SUAS-HIS proposed scheme, in terms of DR at different type of attacks and different $T_S$.

**PDR:** Figure 11 demonstrates the relationship between PDR, number of UAVs, and Simulation times. With 200 active UAVs, the PDR is relatively low for three the SFA, BRUIDS, and Cyber Security System. The reason for this is the fact that in such conditions, some packets fail to reach destination in the designated timeframe. However, since increasing the number of UAVs results in more packets being delivered to the destination, using more UAVs would slightly improve the PDR. When the number of employed UAVs is 200 and 400, due to random factors occurring in simulation, a slight degradation is witnessed in the SUAS-HIS ratio of packet delivery. However, when the simulation times is in range of 600 to 1400, SUAS-HIS is capable of generally outperforming all current approaches. As shown in the Figure 11(a), (b), (c) and (d), SUAS-HIS increases the PDR by more than 18.5, 21.7, and 23.81% those of SFA, BRUIDS, and Cyber Security System models, respectively.

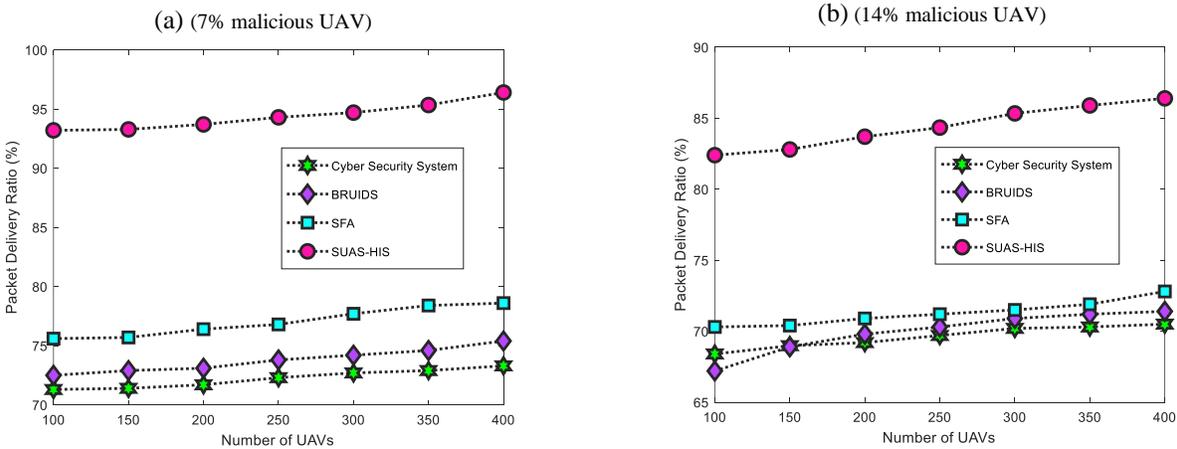

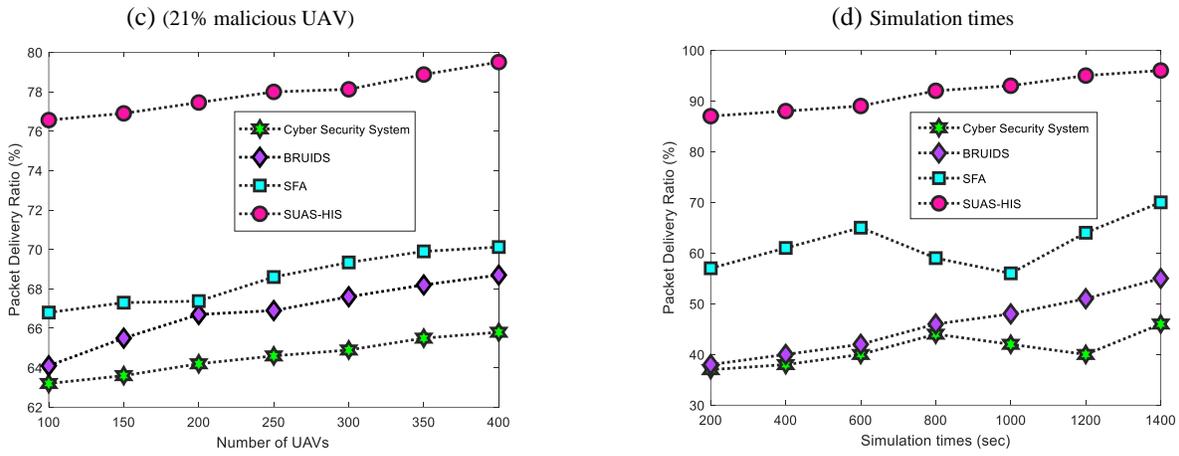

Figure 11: Comparison of the SUAS-HIS proposed scheme, Cyber Security System, BRUIDS and SFA models in term of PDR.

The comparison results of the SUAS-HIS proposed scheme, in terms of PDR at different type of attacks and different $T_S$ are provided in Figure 12.

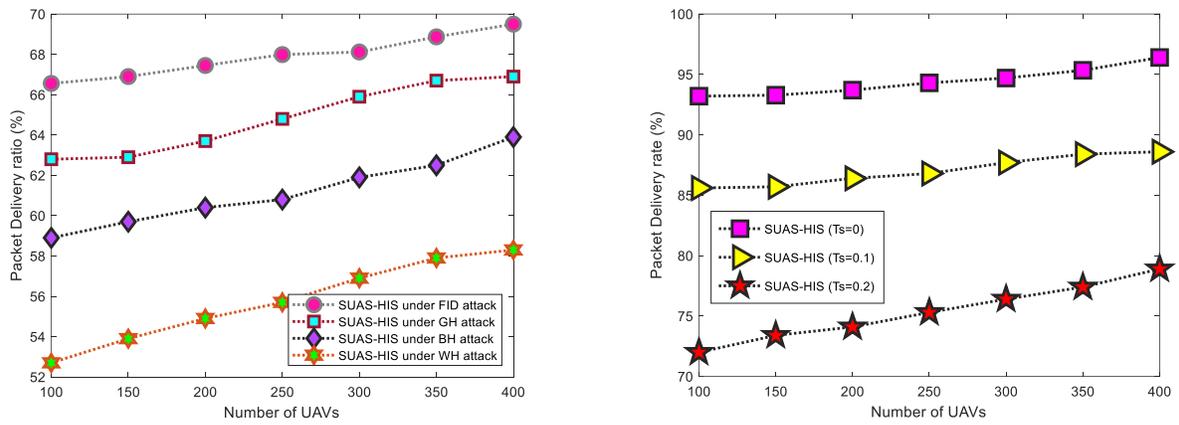

Figure 12: Comparison of the SUAS-HIS proposed scheme, in terms of PDR at different type of attacks and different $T_S$

The simulation was once again considered with a dynamic $T_S$ of 0, 0.1, and 0.2 where the diagrams were obtained using different values of $T_S$. As seen in the diagrams, the performance of the proposed method in the mentioned criteria is presented by increasing the value of parameter $T_S$. When $T_S$ is set to 0, the proposed defense mechanism (SUAS-HIS) gets activated immediately and will react to the operation of malicious UAVs in order to stop them and prevent lots of packets from being removed by the malicious UAVs. If $T_S$ is set to 0.1, SUAS-HIS will be activated after 10 percent of the packets are deleted and the malicious operation will be prevented. Finally, if $T_S$ is set to 0.2, it means that after 20 percent of the packets are lost, SUAS-HIS will start operating.

As is clearly demonstrated in simulation results, employing UAVs to establish a real-time communication is not highly efficient. Instead, they should be potentially employed as assistive mediums to create or improve the level of communication in environments with special characteristics.

For instance, in Linear Sensor Networks based on UAVs or temporary communications, UAVs are an easy-to-deploy and low-cost means of data transfer.

# 6 Conclusion

In a UAS, in order to protect the continuity of the operation, it is imperative to detect attackers, whilst limiting the ratios of FPR and FNR. In this paper, an IDS has been proposed in the SUAS-HIS method to protect against the security problems using the HIS. In the SUAS-HIS, the malicious UAV is robust against four lethal attacks (WH, BH, GH, and FID) so that intrusive operations are quickly identified and removed from the spying missions or top-secret information surveillance. The IDSs are used to detect and respond to attempts to compromise the target system. This design is inspired by HIS. In the mapping, insecure signals are equivalent to an antigen that is detected by antibody-based training patterns and removed from the operation cycle. Among the main uses of the proposed design are the quick detection of intrusive signals and quarantining their activity. SUAS-HIS is more efficient than the Cyber Security System, BRUIDS and SFA approaches under security threats because not only it detects the malicious node beforehand but also it isolates the malicious node and restore the accused node after the penalty period. In our analysis, the performance of our proposed SUAS-HIS scheme is evaluated using NS-3. The results confirmed that our scheme is capable of exhibiting high-levels of security and high ratio of detection (exceeding 92.93%). It addition, our proposed scheme has high PDR (more than 64.41%), low FPR (less than 6.89%), and low FNR (less than 3.95%), in comparison with the other approaches currently being employed.

# Reference

1. Colomina, I. and P. Molina, *Unmanned aerial systems for photogrammetry and remote sensing: A review.* ISPRS Journal of photogrammetry and remote sensing, 2014. 92: p. 79-97.
2. Jamali, S. and R. Fotohi, *DAWA: Defending against wormhole attack in MANETs by using fuzzy logic and artificial immune system.* 2017. 73(12): p. 5173-5196.
3. Ouédraogo, M.M., et al., *The evaluation of unmanned aerial system-based photogrammetry and terrestrial laser scanning to generate DEMs of agricultural watersheds.* Geomorphology, 2014. 214: p. 339-355.
4. Chen, T.M., et al., *Intrusion detection.* IET Engineering & Technology Reference, 2014: p. 1-9.
5. Loukas, G., et al., *A taxonomy and survey of cyber-physical intrusion detection approaches for vehicles.* Ad Hoc Networks, 2019. 84: p. 124-147.
6. García-Magariño, I., et al., *Security in networks of unmanned aerial vehicles for surveillance with an agent-based approach inspired by the principles of blockchain.* Ad Hoc Networks, 2019. 86: p. 72-82.
7. Gurung, S. and S. Chauhan, *A novel approach for mitigating gray hole attack in MANET.* 2018. 24(2): p. 565-579.
8. Panda, S. *GPS Hash Table Based Location Identifier Algorithm for Security and Integrity Against Vampire Attacks*. in *Cyber Security: Proceedings of CSI 2015*. 2018. Springer.
9. Javaid, A.Y., et al. *Cyber security threat analysis and modeling of an unmanned aerial vehicle system*. in *Homeland Security (HST), 2012 IEEE Conference on Technologies for*. 2012. IEEE.
10. Sedjelmaci, H. and S. Senouci, *Cyber security methods for aerial vehicle networks: taxonomy, challenges and solution.* 2018: p. 1-17.
11. Mitchell, R., R. Chen, Man, and C. Systems, *Adaptive intrusion detection of malicious unmanned air vehicles using behavior rule specifications.* 2014. 44(5): p. 593-604.
12. Abusitta, A., et al., *A deep learning approach for proactive multi-cloud cooperative intrusion detection system.* Future Generation Computer Systems, 2019.
13. Rani, C., et al., *Security of unmanned aerial vehicle systems against cyber-physical attacks.* 2016. 13(3): p. 331-342.
14. Sedjelmaci, H., S.M. Senouci, and M.-A. Messous. *How to detect cyber-attacks in unmanned aerial vehicles network?* in *Global Communications Conference (GLOBECOM), 2016 IEEE*. 2016. IEEE.

15. Brust, M.R., et al. *Defending Against Intrusion of Malicious UAVs with Networked UAV Defense Swarms*. in *Local Computer Networks Workshops (LCN Workshops), 2017 IEEE 42nd Conference on*. 2017. IEEE.
16. Yoon, K., et al. *Security authentication system using encrypted channel on UAV network*. in *Robotic Computing (IRC), IEEE International Conference on*. 2017. IEEE.
17. Sedjelmaci, H., et al., *A hierarchical detection and response system to enhance security against lethal cyber-attacks in UAV networks.* 2018. 48(9): p. 1594-1606.
18. Gao, C., et al., *A self-organized search and attack algorithm for multiple unmanned aerial vehicles.* 2016. 54: p. 229-240.
19. Wang, X., et al., *Edge-based target detection for unmanned aerial vehicles using competitive Bird Swarm Algorithm.* 2018. 78: p. 708-720.
20. Seyedi, B., & Fotohi, R. NIASHPT: a novel intelligent agent-based strategy using hello packet table (HPT) function for trust Internet of Things. The Journal of Supercomputing, 1-24. doi:10.1007/s11227-019-03143-7
21. Sarkohaki, F., et al., *An efficient routing protocol in mobile ad-hoc networks by using artificial immune system.* 2017.
22. De Castro, L.N. and F. Von Zuben, Dezembro de, Tech. Rep, *Artificial immune systems: Part I–basic theory and applications.* 1999. 210(1).
23. Mabodi, K., Yusefi, M., Zandiyan, S., Irankhah, L., & Fotohi, R. Multi-level trust-based intelligence schema for securing of internet of things (IoT) against security threats using cryptographic authentication. The Journal of Supercomputing, 1-25. doi:10.1007/s11227-019-03137-5
24. Irimia, R.-E. and M. Gottschling, *Taxonomic revision of Rochefortia Sw.(Ehretiaceae, Boraginales).* 2016(4).
25. Dasgupta, D. and S. Forrest, *Tool breakage detection in milling operations using a negative-selection algorithm.* 1995, Technical report CS95-5, Department of computer science, University of New ….
26. Hatata, A., et al., *An optimization method for sizing a solar/wind/battery hybrid power system based on the artificial immune system.* 2018. 27: p. 83-93.
27. Fotohi, R., & Bari, S. F. (2020). A novel countermeasure technique to protect WSN against denial-of-sleep attacks using firefly and Hopfield neural network (HNN) algorithms. The Journal of Supercomputing, 1-27. doi: 10.1007/s11227-019-03131-x
28. Behzad, S., R. Fotohi, and S. Jamali, *Improvement over the OLSR routing protocol in mobile Ad Hoc networks by eliminating the unnecessary loops.* International Journal of Information Technology and Computer Science (IJITCS), 2013. 5(6): p. 2013.
29. Fotohi, R., et al., *An Improvement over AODV routing protocol by limiting visited hop count.* International Journal of Information Technology and Computer Science (IJITCS), 2013. 5(9): p. 87-93.
30. Fotohi, R., Y. Ebazadeh, and M.S. Geshlag, *A new approach for improvement security against DoS attacks in vehicular ad-hoc network.* International Journal of Advanced Computer Science and Applications, 2016. 7(7): p. 10-16.
31. Consortium, N.-. *ns-3 network simulator*. 2018.
32. Behzad, S., R. Fotohi, and F. Dadgar, *Defense against the attacks of the black hole, gray hole and wormhole in MANETs based on RTT and PFT.* International Journal of Computer Science and Network Solutions (IJCSNS), 2015. 3(3): p. 89-103.
33. Jamali, S. and R. Fotohi, *Defending against wormhole attack in MANET using an artificial immune system.* 2016. 21(2): p. 79-100.
34. Fotohi, R. and S. Jamali, *A comprehensive study on defence against wormhole attack methods in mobile Ad hoc networks.* International journal of Computer Science & Network Solutions, 2014. 2: p. 37-56.
35. Fotohi, R., R. Heydari, and S. Jamali, *A Hybrid routing method for mobile ad-hoc networks.* Journal of Advances in Computer Research, 2016. 7(3): p. 93-103.
36. Manesh, M.R. and N. Kaabouch, *Cyber Attacks on Unmanned Aerial System Networks: Detection, Countermeasure, and Future Research Directions.* Computers & Security, 2019.

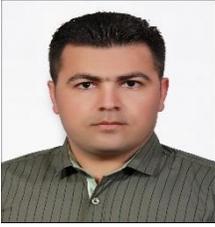 **Reza Fotohi** received his B.Sc. degree in computer software technology engineering from university of applied science and technology, Shabestar, IRAN (2009) (selected as the best student). M.Sc. degree in computer software engineering from Islamic Azad University (IAU), Shabestar, IRAN (2013), under the supervision of Dr. Shahram Jamali associate professor University of Mohaghegh Ardabili (UMA) (selected as the best student). From 2012 until now, he is a lecturer in the department of computer engineering, PNU and UAST university. Iran. From 2015, M.Sc. Fotohi is a member of reviewer the Applied Soft Computing (Elsevier, ISI-JCR), Artificial Intelligence Review (Springer, ISI-JCR), Human-centric Computing and Information Sciences (Springer, ISI-JCR), The Journal of Supercomputing (Springer, ISI-JCR), Wireless Personal Communications Journal (Springer, ISI-JCR), National Academy Science Letters (Springer, ISI-JCR), KSII Transactions on Internet and Information Systems (ISI-JCR), and The Turkish Journal of Electrical Engineering & Computer Sciences (ISI-JCR). His research interests include computer networks, Internet of Things (IoT), Unmanned Aerial Systems (UASs), wireless networks, network security, cyber security, Artificial Immune System (AIS), Fuzzy Logic Systems (FLS), and NS-2, NS-3, Cooja Simulation. He is currently student member of the IEEE. He is author and coauthor of more than 10 journal and international conferences papers. His papers have more than 175 citations with 9 h-index.